\documentclass[a4paper, amsfonts, amssymb, amsmath, reprint, showkeys, nofootinbib, twoside]{revtex4-1}
\usepackage[english]{babel}
\usepackage[utf8]{inputenc}
\usepackage[colorinlistoftodos, color=green!40, prependcaption]{todonotes}
\usepackage[pdftex, pdftitle={Article}, pdfauthor={Author}]{hyperref} 
\begin{document}
\title{On the viability of Locked Inflation}
\author{Michael Zantedeschi}
    \email[]{michael.zantedeschi1@physik.uni-muenchen.de}
    \affiliation{Arnold Sommerfeld Center, Ludwig-Maximilians-Universität, Theresienstraße 37, 80333 München, Germany,\\
Max-Planck-Institute für Physik,  Föhringer Ring 6, 80805 München, Germany
}

\date{\today} 
\begin{abstract}
We review locked inflation and we critically address phenomenological and consistency issues that have appeared in the literature. A natural window of opportunity is found for the original scenario. Moreover, a simple way to enlarge the parameter space is proposed by considering an equivalently natural structure for the locking field. Finally, an estimate of the amount of topological defects at the end of inflation is given; due to the low inflationary scale, it turns out to be negligible. 
\end{abstract}
\maketitle
\section*{Introduction}
According to the standard lore, slow-roll conditions are imposed on the inflaton in order to preserve the equation of state $\rho\approx -p$. As a consequence, the curvature of the inflaton in the potential $m^2$ is much smaller than the inflationary scale fixed by the Hubble parameter $H^2$. In supergravity-inspired inflationary models, such a condition cannot be fulfilled since supersymmetric flat directions receive corrections proportional to the inflationary scale \cite{gia95}. This is known as $\eta$ problem. One of the most minimal solution to this problem is locked inflation \cite{dvali03,dvalikachru2}. In Sec. I this two-field fast-roll scenario is analyzed and reviewed. In this model the inflationary field is locked in a saddle point via a locking field which is oscillating and redshifted due to the gravitational interaction. The false vacuum energy in this configuration sources an exponential expansion. Different phenomenological and consistency constraints in the literature reduced the parameter space of such a model \cite{schalm04} to the point where it was claimed to be fully ruled out \cite{copeland}. In Sec. II, we review these claims and we explain why this need not be the case. In Sec III, we summarize our results.

\section{The model}
Consider the following potential of two weakly coupled fields 
\begin{equation}
    V(\Phi,\phi)= M_\Phi^2 \Phi^2 + \lambda \Phi^2 \phi^2 + \alpha \left( \phi^2 - M_{\star}^2\right)^2,
\end{equation}
with $\alpha \sim M^4/M_p^4$, $M_\Phi^2\sim M^4/M_p^2$ and $M_{\star}\sim M_p$, $\lambda \sim 1$. $M$ here is some intermediate supersymmetry (SUSY) breaking scale which we take of order TeV and we are ignoring numerical factors of order one. This potential has two stationary points: one at $\phi=\Phi=0$ and a true vacuum at $\phi=M_{\star}$, $\Phi=0$. We assume that the initial configuration starts in the saddle point and that at the beginning $\Phi_0 \gg \alpha M_\star/\lambda$, but such that the energy content of the universe is dominated by the false vacuum energy $\rho_0 = \alpha M_\star^4=H^2M_p^2$. In this situation, it follows that $M_\Phi \geq H$ and the locking field will evolve according to $\Phi(t)\simeq \Phi_0 e^{-3/2Ht}\cos{M_\Phi t}$. \\
Via the $\lambda$ coupling, an effective mass term will be generated for the inflaton field $\phi$, hence realizing the locking
\begin{equation}
\label{effmass}
    m_{\phi}^2(t)= \langle \Phi^2\rangle (t) - \alpha M_\star^2,
\end{equation}
where the average is performed over the sinusoidal oscillations. Such an approximation is valid as long as the locking field crosses zero fast enough so that a restoration of the $\phi$ true minimum happens on a time window which is smaller than $1/m_\phi$. This leads to the condition
\begin{equation}
    \Delta t \sim \Delta \Phi\, e^{3/2\,Ht}\frac{1}{\Phi_0M_\Phi}\leq \frac{1}{m_\phi} .
\end{equation}
Violation of the inequality gives a number of e-folds $dN=d\ln{a}$, $a$ being the scale factor, which is 
\begin{equation}
\label{oscillconstr}
    N \sim -\frac{1}{3}\ln{\alpha}\sim 50.
\end{equation}
This corresponds to the moment where the averaging over the oscillation effectively breaks. \\
Another natural way to evaluate the number of e-folds is the moment where the effective mass in \eqref{effmass} becomes negative. This gives
\begin{equation}
\label{tachyconstr}
    N\sim \frac{1}{3}\ln{\frac{M_\star^2}{M_\Phi^2}} \sim 50.
\end{equation}
Both \eqref{tachyconstr} and \eqref{oscillconstr} should be taken into account when evaluating the number of e-folds. As a general criterion, the condition breaking first will mark the end of inflation. A nice way to understand the two constraints graphically is given in fig. $1$. 
\begin{figure}[h]
\centering
\includegraphics[width=0.26\textwidth]{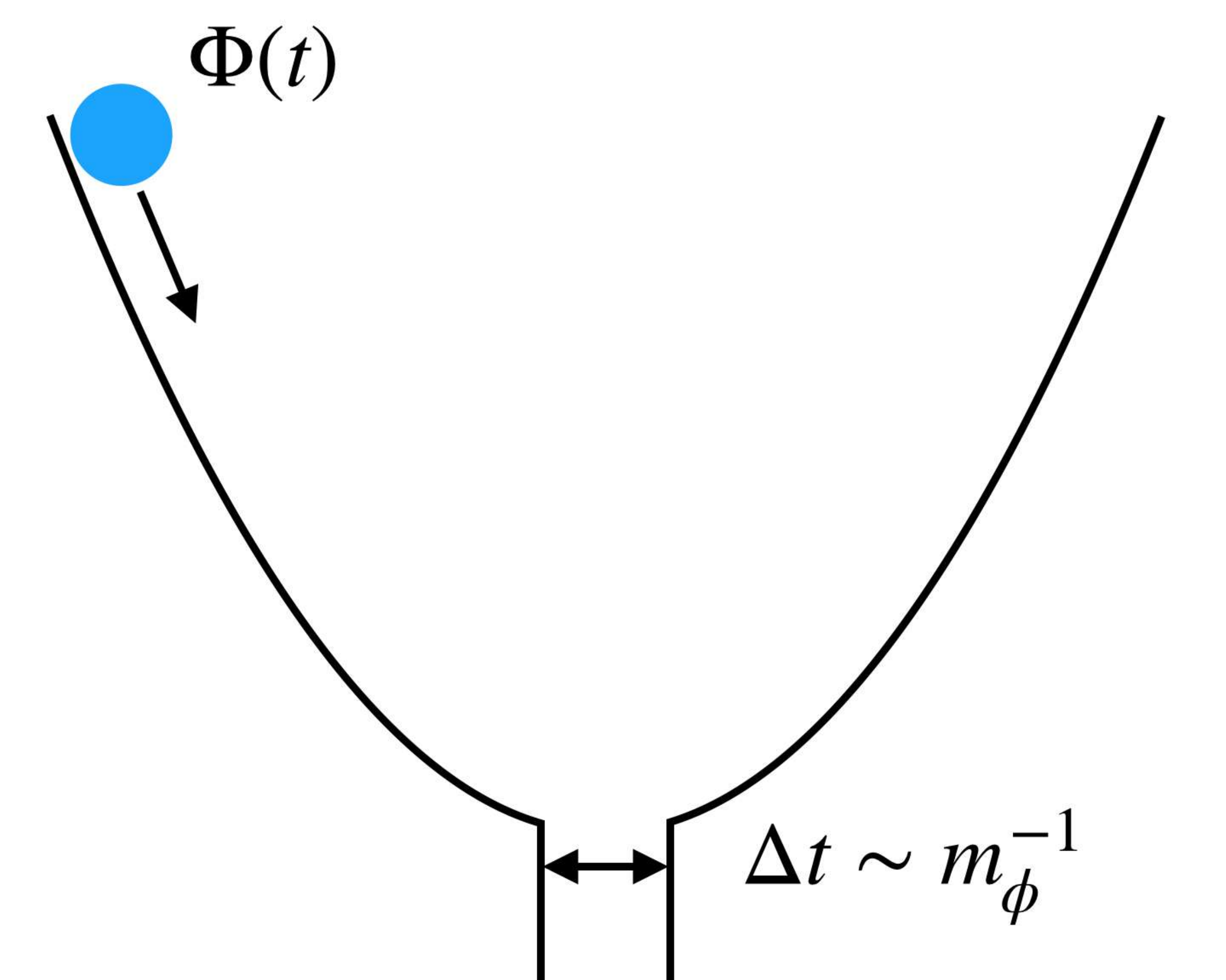}
\caption{In order for the ``pinball'' $\Phi$ to not finish in the hole and trigger a phase transition, two conditions must be met. Firstly, the ball should pass through the hole fast enough, and secondly, the amplitude of the ball oscillation, which is redshifted via gravitational friction, should be bigger than the hole size.}
\end{figure}

Before moving further, we would like to briefly address the phenomenological status of this model. In inflationary models $50$ e-folds are enough for CMB measurement, however they are not enough in order to solve the horizon and homogeneity problem. Usually a number of e-folds for this to happen is $\sim 60$. This is solved if we look at the pre-locked inflationary period. Mainly we are interested in two facts: 1) the initial patch is of order $1/H$, 2) the locking field has initial value different from zero.\\ 
The idea proposed in \cite{dvali03} is to use a bubble mechanism in the spirit of Guth's inflationary model \cite{guth81} in order to produce the necessary initial conditions. Imagine to add a false minimum in the $\Phi$'s potential at roughly $M_p$ separated from the other minimum by a barrier and consider the initial condition $\Phi \sim M_p$ and $\phi\sim0$. Because the minimum is added at $M_p$, such a modification of the potential does not alter the considerations previously made for locked inflation. Notice also that in this configuration, due to the $\lambda$ coupling, the effective mass for $\phi$ is Planckian and can be ignored in the dynamics. Effectively, we are left with the $\Phi$ field sitting in a false vacuum, generating a de Sitter expansion (we can choose our potential so that the Hubble parameter $H$ is the same order as the one in locked inflation, but in principle different choices are possible). Via the bubble mechanism a transition to the true minimum can take place and the described locked inflation takes place. \\
The probability of a transition to the true vacuum is given by the Hawking-Moss formula \cite{hawk82} for the instanton
\begin{equation}
\label{hawkmoss}
P\propto \exp{\left[\frac{3\,M_p^4}{8}\left(\frac{1}{V_{fv}}- \frac{1}{V_{max}} \right)\right]} ,   
\end{equation}
up to subexponential prefactors. Here $V_{fv}$ and $V_{max}$ are the value of the potential in the false vacuum potential and on top of the barrier respectively. Such a formula can be understood in terms of stochastic approach to inflation \cite{staro88}. The inflationary field can be written as
\begin{equation}
    \Phi(x)= \overline{\Phi} + \hat{\Phi},
\end{equation}
where the first term is the classical field outside the horizon where the fluctuations are frozen, while the second term describes the field within the horizon whose modes satisfy the classical equation of motion. The idea is that upon horizon crossing, the modes coming from $\hat{\Phi}$ will freeze with different phases acting as white noise in the equation of motion of the classical slow-rolling field $\overline{\Phi}$. Under these assumptions, one can derive a Fokker-Planck equation for the probability distribution $P(\Phi,t)$
\begin{equation}
    \frac{\partial P}{\partial t}= \frac{H^3}{8\pi^2}\frac{\partial^2 P}{\partial \Phi^2}+ \frac{\partial }{\partial \Phi}\left( \frac{P}{3H}\frac{dV}{d\Phi}\right).
\end{equation}
Assuming the quasi-stationarity of the process, the asymptotic solution to the above yields the Hawking-Moss result \eqref{hawkmoss}. Indeed, it follows that the starting patch for locked inflation must be of the order of the brownian motion steps, which is $1/H$, and the locking field $\Phi$ starts its motion from the top of the barrier. \\
Therefore, both the horizon and the flatness problems are solved. In fact, the observed patch becomes $\frac{1}{H}e^N\frac{T_r}{T_{today}}\sim 10^{37}cm \gg 1/H_{today}$. Regarding the homogeneity problem, it can be argued that it is partially solved by the previous de Sitter expansion ante bubble nucleation. 

\section{Constraints on the parameter space of locked inflation}
There are mainly three problems affecting the parameters space of locked inflation.
\begin{itemize}
    \item The transition of $\phi$ at the end of inflation to its true vacuum might generate a period of extra inflation baptized saddle inflation in \cite{schalm04}. This inflationary stage is of the slow-roll type. There are three possibilities: 1) this slow-roll phase lasts longer than fifty e-folds, thus washing out all possible imprints of locked inflation; 2) it lasts less than fifty e-folds. However it is shown by the authors of \cite{schalm04} that this would produce an unacceptable number of very massive black holes within our horizon or 3) saddle inflation does not take place because of the choice of our parameters. This is the phenomenological viable situation we will assume.
    \item  Radiative corrections pointed out in \cite{dvalikachru2} and \cite{copeland} modify the potential at one loop and might move the vacuum of $\Phi$ too far from zero. In this case, the true vacuum of the inflaton field will never be reached and inflation will last forever.
    \item Parametric resonances, indicated in \cite{dvalikachru2} and expanded in \cite{copeland}, might ruin inflation within the first e-folds. In particular, this lead to claim that the parameter space of locked inflation is fully ruled out \cite{copeland}.
\end{itemize}
Regarding the last point we disagree for three reasons. First of all, as discussed below, a window of opportunity within their arguments can still be found. Secondly, the performed numerical study, as already pointed in the aforementioned article, does not fully take into account non-linearities. In this sense, it is not considered how the backreaction of the locking field affects the model. It is not clear whether such non-linearities drive the system out of the parametric resonant trajectories. Thirdly, simply considering a complex scalar field as a locking field, fully removes the constraints coming from parametric resonances. In what follows we are going to make these statements more clear.

\subsection{Constraints}
As already pointed out in \cite{dvalikachru2}, the potential gets loop corrections of the Coleman-Weinberg type due to the inflaton breaking supersymmetry. These read (at one-loop order)
\begin{equation}
    \Delta V = \frac{m_\phi^2}{64\pi^2}\Phi^2\ln{\left(\frac{\Phi^2}{\mu^2} \right)},
\end{equation}
where $\mu$ is the renormalization scale to be fixed. This type of correction to the Kähler potential, arising from $\phi$ loop, does not cancel in a supersymmetric scenario even though all other corrections are absent \cite{dvali94}. Such a contribution might shift the minimum of $\Phi$ from zero to a value bigger than $\Phi_c$, $\Phi_c$ being the critical value of $\Phi$ for which $m^2_\phi$ becomes zero. If this was the case, locked inflation would last forever.\\
To fix $\mu$ we imposed the on-shell condition
\[
V_{,\Phi\Phi}\left(\Phi=\Phi_c \right)= M_\Phi^2,
\]
then, requiring the minimum to be at a value smaller than $\Phi_c$ i.e. $V_{,\Phi}\left(\Phi_c \right)\geq 0$ gives the following constraint on the masses
\begin{equation}
\label{1loopconstr}
    M_\Phi^2\geq \frac{m_\phi^2}{16\pi^2}\sim 10^{-2}m_\phi^2.
\end{equation}
Notice that even saturating the bound would give a negligible change in the number of e-folds in \eqref{tachyconstr} i.e. $\Delta N /N\ll 1$.\\
Next we will turn to parametric resonances (see \cite{kofman97} for the notions utilized in the following).  
As in \cite{copeland} the equation of motion for the inflationary field $\phi$ is 
\begin{equation}
    \Ddot{\phi}+ 3H\Dot{\phi}+\left(\lambda \Phi(t)^2-m_\phi^2 \right)\phi=0,
\end{equation}
which can be rewritten as
\begin{equation}
\label{mathieu}
    \chi'' + \left(2q(\tau)(1-\cos{2\tau})-b \right)\chi=0,
\end{equation}
where 
\[\tau \doteq m_\phi t,\quad \chi\doteq \exp{\left(3/2Ht\right)}\phi,\quad b\doteq \frac{m_\phi^2}{M_\Phi^2}+\frac{9}{4}\frac{H^2}{M_\Phi^2},\]\[\quad q(\tau)\doteq q_0\exp{(-3/2Ht)},\quad q_0\doteq \frac{\Phi_0^2}{4M_\Phi^2}\] and the prime denotes derivative with respect to $\tau$. Note that from the constraint \eqref{1loopconstr} it follows $b\leq 100$.\\
The above eq. \eqref{mathieu} is known as Mathieu's equation. We assume the factor $q(\tau)$ to be almost constant, implying $H\leq 10^{-1} \,M_\Phi$. Then a general solution to the equation of motion is given by 
\begin{equation}
    \chi(\tau)=e^{s\tau}f(\tau),
\end{equation}
where $f(\tau)$ is a function periodic in $\tau$ and $s$ is known as Floquet exponent. Such a solution has a clear physical interpretation. Whenever the sinusoidal term in \eqref{mathieu} approaches zero, a violation of the adiabaticity condition $|\Dot{\omega}|/\omega^2\geq 1$ leads to an uncontrolled production of particles. This is nothing but a broad resonance with cutoff mode $k_{max}\sim \Phi_0$ that takes place at each oscillation for a time interval $\Delta t_\star\sim k_{max}^{-1}$. The average value of $s$ can be calculated both numerically \cite{copeland} and analytically \cite{kofman97} and turns out to be $\overline{s}\sim 0.1$. In turn, such a production of particles can be redshifted enough by the inflationary expansion as the original field $\phi$ scales as
\begin{equation}
    \phi(t) \propto e^{\left(\overline{s}-\frac{3H}{2M_\Phi}\right)\tau}.
\end{equation}
Combining this with the assumption that $q(\tau)$ is quasi-constant we get 
\begin{equation}
\label{redshiftcond}
    H\sim 10^{-1}M_\Phi.
\end{equation}
From this point on, the analysis in \cite{copeland} proceeds by integrating over the enhanced homogeneous modes to obtain $\langle\phi^2\rangle$ and it is found that within one e-fold, such a term appearing in the background equation for $\Phi$ (as an effective mass term via $\lambda$ coupling) is much bigger than the mass $M_\Phi$. It is thus claimed that the locking field oscillatory behaviour is no longer reliable. However, it could very well be that a change in the effective mass of the background slows its oscillatory frequency, halting the resonance in the $\phi$ modes. We are not interested in pursuing explicitly this, since the restrictions coming from parametric resonances are fully lifted by simply considering a more complicated field structure for the locking field. In fact, already a complex scalar field removes the oscillatory term in \eqref{mathieu} because different components of the field will cross the zero value at different moments. Such a solution not only greatly increases the parameter space of locked inflation, but it also removes the need of an oscillatory averaging in the $\lambda$ coupling and the constraint \eqref{oscillconstr} on the number of e-folds can be dropped.\\ 
The next consistency issue was proposed in \cite{schalm04}. At the end of locked inflation, the classical field $\phi$ is ready to roll down to its true vacuum at $\phi=M_\star$. Neglecting $\Phi$, which by now has been fully redshifted, an explicit solution to the equation of motion for $\phi$ can be found if we ignore the quartic interaction in the potential. In this regime, we are effectively describing a slow-roll down the hill inflationary period. Hence, one can find a bound on the value of $\phi$'s VEV, $M_\star$, by requiring this slow roll period to last less than one e-fold. Practically this is done by introducing a slow roll parameter $\epsilon \sim \Dot{\phi}^2/V$. The end of the slow roll is obtained when $\epsilon$ becomes of order one. This in turn gives the number of e-folds by matching with the solution found in this particular approximation. Requiring the number of e-folds to be smaller than one gives \cite{schalm04}
\begin{equation}
\label{Nsaddle}
    N_f = \frac{2}{3(\delta -1)}\ln{\frac{2M_p\delta}{3(\delta+1)^2 \phi_0}}\leq 1,
\end{equation}
where $\delta =\sqrt{1+\frac{4}{3}\frac{M_p^2}{M_\star^2}}$ and $\phi_0$ is the initial value of the field $\phi$ (the velocity $\Dot{\phi_0}=0$ is assumed to be zero). In particular it is assumed that $\phi_0\sim H$ because of fluctuations in both \cite{schalm04} and \cite{copeland}. However as one can easily infere from \eqref{mathieu}, locked inflation will end roughly when $q(\tau)\sim b$ resonantly. Hence, in principle, different initial conditions should be implemented to solve the equation of motion for $\phi$. It turns out that such corrections due to resonances are negligible. This is because at the end of locked inflation, the energy stored in the oscillatory field $\Phi$ is roughly $\rho\sim M_\Phi^2m_\phi^2$. Even assuming that the totality of such an energy is converted into kinetic energy via $\Dot{\phi}_0^2$, a negligible change to \eqref{Nsaddle} is obtained. The constraint coming from saddle inflation is then the same given by \cite{schalm04}
\begin{equation}
\label{uppervev}
    M_\star \leq 10^{-2}M_p.
\end{equation}
A lower bound on $M_\star$ comes from the reheating temperature. As already pointed out in \cite{dvali03} locked inflation can have a reheating temperature up to the TeV order. This can be immediately seen since $T_r\sim \sqrt{M_p\, H}$. With such a low inflationary scale it is not possible to produce the desired amount of density perturbations. Such a problem was addressed already in \cite{dvali03} and the solution is to use the mechanism of modulated reheating \cite{dvali02}. The idea is that reheating happens via an effective Yukawa-like coupling modulated by another field called modulating field. Fluctuations in such a field lead to fluctuations in the imprint of density perturbations. Hence $O(\text{TeV})$ reheating, or even lower, becomes possible. The lowest possible value, however, is fixed by the requirement that neutrino decoupling should take place. Imposing $T_r\geq 10^{-2}\text{GeV}$ yields $M_\star \geq 10^{-5}M_p$. Notice, however, that we are mainly interested in the upper bound \eqref{uppervev} being the one giving a negligible change in the number of e-folds. \\
Finally, a bound on the parameter space of locked inflation could come from the production of topological defects due to the structure of the vacuum-manifold of the inflationary field $\phi$ when it relaxes to its true VEV. In the original model, $\phi$ has a $\mathbf{Z}^2$ symmetry and hence would produce domain walls. The following estimate equally applies if one considers more complicated structures such as strings or monopoles. Due to the low inflationary scale, it is not possible to use the Kibble mechanism \cite{kibble76} to estimate the correlation length for the production of topological defects. We compute such a quantity by estimating the criticality length fixed by the adiabaticity violation at the criticality time, i.e. we use
\begin{equation}
\label{adiabat}
   {\frac{|\dot{\omega}_k|}{\omega^2} }{\big|_{t=t_c}}\geq 1, \quad \omega_k(t)^2 = k^2 + \langle \Phi^2\rangle(t) - \alpha M_{\star}^2.
\end{equation}
Saturating \eqref{adiabat} at the critical moment $t_c$ (the moment when the minimum of the inflationary field is restored) will give a critical length which we interpretate as a correlator length. Hence the number density of produced topological defects is
\begin{eqnarray}
\label{topdef}
    n \sim \alpha M_\star^2 H\sim 10^{-30}\text{GeV}^3,
\end{eqnarray}
which is negligible. Such a result could have been already foreseen because of the low inflationary scale of the model ($H\sim O(10^{-3}\text{eV}))$ and it is reflected by the presence of the weak coupling $\alpha$ in \eqref{topdef}. Hence no real constraint comes from the production of topological defects.

\section{Conclusions and Outlook}
In this article the steps that lead to the conclusions of \cite{copeland} that the parameter space of locked inflation is fully ruled out are critically addressed. It is found that a window of opportunity for the model is still open for $M_\star\leq 10^{-2}M_p$ and $M_\Phi\sim 10^{-1}H$. It should be stressed that the latter condition can be fully dropped considering a complex scalar field instead of a real one as a locking field. This is by no means not a complication for the model since such a requirement is natural for supersymmetric potentials. Hence, an elegant and viable solution to the $\eta$ problem is proposed without requiring any fine-tuning.\\
Finally, the production of topological defects due to the inflationary field relaxation to its true vacuum is taken into consideration. An estimate based on the adiabaticity condition led us to the conclusion that no relevant amount of topological defects is produced.

\paragraph*{Acknowledgments}
I am grateful to Gia Dvali for useful inspiration and feedback on this work and to Georgios Karananas for carefully reading the manuscript.
\bibliography{citations1.bib}
\end{document}